\documentclass[aps,prl,twocolumn,superscriptaddress]{revtex4}
\usepackage{graphicx}

\newcommand{\E}{{\rm e}}
\newcommand{\I}{{\rm i}}

\begin{document}

% Title of the article
\title{Zero-Phonon Line Broadening and Satellite Peaks in Nanowire Quantum Dots: The Role of Piezoelectric Coupling}

% Abbreviated title for the page headers
%\titlerunning{Piezoelectric Electron-Phonon Coupling in Nanowire QDs}

% Authors
\author{Carsten Weber}
\email[]{carsten.weber@teorfys.lu.se}
\affiliation{Mathematical Physics, Lund University, Box 118, SE-221 00, Lund, Sweden}
\author{Greta Lindwall}
\affiliation{Mathematical Physics, Lund University, Box 118, SE-221 00, Lund, Sweden}
\affiliation{Swerea KIMAB, Drottning Kristinas v\"ag 48, 114 28 Stockholm, Sweden}
\author{Andreas Wacker}
\affiliation{Mathematical Physics, Lund University, Box 118, SE-221 00, Lund, Sweden}

% \author{%
%   Carsten Weber\textsuperscript{\textsf{\bfseries 1,\Ast}},
%   Greta Lindwall\textsuperscript{\textsf{\bfseries 1,2}},
%   Andreas Wacker\textsuperscript{\textsf{\bfseries 1}}}
% 
% % Abbreviated list of authors for the page haeders
% \authorrunning{C. Weber, G. Lindwall, A. Wacker}

%E-mail-address of corresponding author
% \mail{e-mail
%   \textsf{carsten.weber@teorfys.lu.se}, Phone
%   +46-46-222 4458, Fax +46-46-222 4416}

% author's affiliations/addresses
% \institute{%
%   \textsuperscript{1}\,Mathematical Physics, Lund University, Box 118, 221 00 Lund, Sweden\\
%   \textsuperscript{2}\,Corrosion and Metals Research Institute, Drottning Kristinas v\"ag 48, 114 28 Stockholm, Sweden}

% \received{XXXX, revised XXXX, accepted XXXX} % do not change, will be filled in by the publisher
% \published{XXXX} % do not change, will be filled in by the publisher
\date{\today, to appear in Physica Status Solidi B}

%Please select four to six PACS-codes from the enclosed list (PACS.txt) or from www.aip.org/pacs)
 % For example: 71.20.Ps

% \abstract{%
% This is a macro for the typesetting of two-column text in an
% abstract. It will typeset the two arguments in \abstcol{}{} as the
% left and right column inside the abstract box. At the
% columnbreak there will be always a columnbreak (\par), so both
% columns start with a new paragraph. No automatic column height
% balancing is done.
%
% If used with a \titlefigure it will silently output both
% parameters as consecutive paragraphs.
%
% The macro is defined exclusively inside the argument of \abstract{};
% if used outside it will raise an error.
%
% Usage: \abstcol{<left column>}{<right column>}
% \abstcol{

\begin{abstract}
We investigate the influence of the one-dimensional character of the phonon modes in a catalytically grown GaAs nanowire on the absorption spectrum of an embedded quantum dot, focusing on the contribution from the piezoelectric coupling. In general, the reduced dimensionality of the phonons leads to spectral side peaks and a zero-phonon line broadening due to the energetically lowest (acoustic) phonon mode. While the deformation potential predominantly couples to radial modes, the piezoelectric interaction can also couple strongly to modes of axial character, leading to additional absorption features. The contribution of the piezoelectric coupling to the zero-phonon line is negligible.
\end{abstract}

% If \titlefigure is given, it takes as its mandatory parameter the
% name (without extension) of some figure file.
%\titlefigure[height=3.1cm]{empty2w}
%\titlefigurecaption{%
%  This is the caption of the \emph{optional} abstract figure. If
%  there is no figure here, the abstract text will fill both columns.}

\maketitle   % please do not remove

\section{Introduction}
Recent epitaxial progress in the fabrication of catalytically grown nanowires as well as the possibility to include heterostructures has led to an increased interest in quantum dots (QDs) in nanowires \cite{Bjork:NanoLett:02,Agarwal:ApplPhysA:06}. QDs are interesting for the study of quantum information and cryptography \cite{Moreau:PhysRevLett:01,Kako:NatureMater:06,Borgstroem:NanoLett:05}, requiring a good control of the dephasing mechanisms. Since the QDs are embedded in a strongly confined solid state matrix, it is important to take into account the influence of the geometry of the system on the phonon modes which offer an important decoherence channel. 

Recently, the influence of the phonons interacting with an electronic two-level system via the deformation potential coupling was investigated in a GaAs wire \cite{Lindwall:PhysRevLett:07}. Here, we extend this analysis to include the interaction with phonons via the piezoelectric coupling which may be of particular relevance for wurtzite nanowires. In comparison with bulk modes which lead to continuous phonon side bands attached to the zero-phonon line (ZPL) \cite{Borri:PhysRevLett:01,Krummheuer:PhysRevB:02,Forstner:PhysStatusSolidiB:02,Forstner:PhysRevLett:03}, the interaction with wire phonons leads to a set of phonon side peaks. These  reflect the multitude of phonon modes arising due to the spatial confinement in the lateral direction. In addition, while bulk phonons do not lead to a finite linewidth of the excitonic transition without resorting to higher order processes \cite{Forstner:PhysStatusSolidiB:02,Zimmermann::02,Muljarov:PhysRevLett:05,Machnikowski:PhysRevLett:06,Muljarov:PhysRevLett:07}, the lowest (acoustic) wire phonon mode leads to a temperature dependent ZPL broadening. In this paper, we focus on the piezoelectric interaction which can lead to additional absorption features as compared to the case of the deformation potential coupling due to different selection rules in the coupling mechanisms.

\section{Theory}

In the following, we give a brief outline of the calculation of the absorption spectrum and the phonon modes following Ref. \cite{Lindwall:PhysRevLett:07} as well as the QD parameters used. The different electron-phonon coupling mechanisms are considered in Section \ref{coup}.\\

\noindent{\bf Hamiltonian}:
% \paragraph{Hamiltonian}
We consider an interband two-level system (conduction level $|c\rangle$, valence level $|v\rangle$) interacting with the wire phonon modes via a diagonal coupling after a $\delta$-pulse excitation by a classical light field. After the initial excitation, the system is described by the Hamiltonian
\begin{equation}
H = H_0 + H_{\rm el-ph};
\end{equation}
the first term,
\begin{equation}
H_0 = \left( \varepsilon_v a^\dagger_v a^{\phantom \dagger}_v + \varepsilon_c a^\dagger_c a^{\phantom \dagger}_c \right) + \sum\limits_{q \kappa} \hbar \omega_{q \kappa} b^\dagger_{q \kappa} b^{\phantom \dagger}_{q \kappa},
\end{equation}
describes the kinetics of the electronic two-level system and the phonons. Here, $a^\dagger_i (a^{\phantom \dagger}_i)$ denotes the creation (annihilation) operator of an electron in state $|i\rangle$ with energy $\varepsilon_i$, while $b^\dagger_{q \kappa} (b^{\phantom \dagger}_{q \kappa})$ denotes the creation (annihilation) operator of a phonon with quasi-momentum $q$ and dispersion $\omega_{q \kappa}$, where the modes are labeled by $\kappa$.
%The electron-light interaction is given in dipole approximation,
%H_{\rm el-light} = E(t) d_{v c} a^\dagger_v a^{\phantom \dagger}_c + {\rm H.c.},
%\end{equation}
%with the interband dipole moment $d_{v c}$ and the projection of the field along the dipole direction $E(t)$. 
The interaction with the phonons is given by \cite{Mahan::00}
\begin{equation}
H_{\rm el-ph} = - a^\dagger_c a^{\phantom \dagger}_c \sum\limits_{q \kappa} \left( g^{q \kappa}_{v v} - g^{q \kappa}_{c c} \right) \left( b^{\phantom \dagger}_{q \kappa} + b^\dagger_{-q \kappa} \right).
\end{equation}
Here, $g^{q \kappa}_{i i} = \int {\rm d}^3x \, \varphi^*_i({\bf x}) V^i_{q \kappa}({\bf x}) \E^{\I q z} \varphi_i({\bf x})$ is the electron-phonon coupling element with the interaction potential\\$V^i({\bf x}) = \sum_{q \kappa} V^i_{q \kappa}({\bf x}) \E^{\I q z} ( b^{\phantom \dagger}_{q \kappa} + b^\dagger_{-q \kappa} )$, and the interaction is taken to be diagonal. No higher order phonon terms are considered here.\\

\noindent{\bf Absorption}:
%\paragraph{Absorption}
The absorption is determined by the imaginary part of the susceptibility via the macroscopic polarization 
$P(t) \sim d_{v c} p(t) + {\rm c.c.}$, where $d_{v c}$ is the interband dipole moment. The microscopic polarization $p(t) = \langle a^\dagger_v a^{\phantom \dagger}_c \rangle$ is calculated via the independent Boson model (IBM) which can be solved analytically for an arbitrary number of phonon modes. The solution for $t>0$ is given by \cite{Krummheuer:PhysRevB:02,Forstner:PhysStatusSolidiB:02,Forstner:PhysRevLett:03,Mahan::00}%,Zimmermann::02}
\begin{eqnarray}
&&p(t) = p(0) \E^{-\I (\omega_{\rm gap} + \Delta) t} \times\\
&&\times \exp\left[ -\int {\rm d}\omega \frac{J(\omega)}{\omega^2} \left( 4 n(\omega)  \sin^2 \frac{\omega t}{2} + 1 - \E^{-\I \omega t} \right) \right], \nonumber
\end{eqnarray}
with the transition energy $\omega_{\rm gap} = \varepsilon_c - \varepsilon_v$, the Bose-Einstein distribution $n(\omega) = [ \exp(\hbar \omega/k_{\rm B} T) - 1 ]^{-1}$  (temperature $T$), and the polaron shift and the phonon spectral density, respectively given by
\begin{eqnarray}
&&\Delta = -\int {\rm d}\omega \frac{J(\omega)}{\omega},\\
%&&\Delta = -\int {\rm d}\omega J(\omega)/\omega,\\
&&J(\omega) = \sum\limits_{q \kappa} \frac{|g^{q \kappa}_{v v} - g^{q \kappa}_{c c}|^2}{\hbar^2} \delta(\omega - \omega_{q \kappa}).
\end{eqnarray}

\noindent{\bf Phonon modes}:
%\paragraph{Phonon modes}
The phonon modes are calculated within an isotropic continuum model based on Ref. \cite{Stroscio:JApplPhys:94}.
% leading to the elastic equation
% \begin{equation}
% \rho_{\rm m} \partial_t^2 {\bf u} = \lambda \nabla^2 {\bf u} + 2 \mu \nabla (\nabla \cdot {\bf u}),
% \end{equation}
% where ${\bf u}({\bf x})$ is the displacement, $\rho_{\rm m}$ the mass density, and $\lambda,\mu$ are the Lam\'e parameters \cite{Lai::93}.
Assuming free-surface boundary conditions, the solutions are uniquely determined, taking the general quantized form
\begin{equation}
u_i({\bf x}) = \frac{1}{\sqrt N} \sum_{q \kappa} u_{q \kappa}^i({\bf x}) \E^{\I q z} ( b^{\phantom \dagger}_{q \kappa} + b^\dagger_{-q \kappa} )
\end{equation}
($i = r,\varphi,z$; $N$: number of unit cells). In the following discussions, we restrict to compressional modes ($u_\varphi = 0$). Torsional modes ($u_\varphi \neq 0$) do not couple via the deformation potential coupling; for the piezoelectric coupling, this is taken to be an approximation. Furthermore, modes with a $\varphi$-dependence (so-called flexural modes) are disregarded, since we assume a QD model which is azimuthally symmetric, in which case the coupling to these modes vanishes.

\begin{figure}[htb]
\includegraphics*[width=\linewidth]{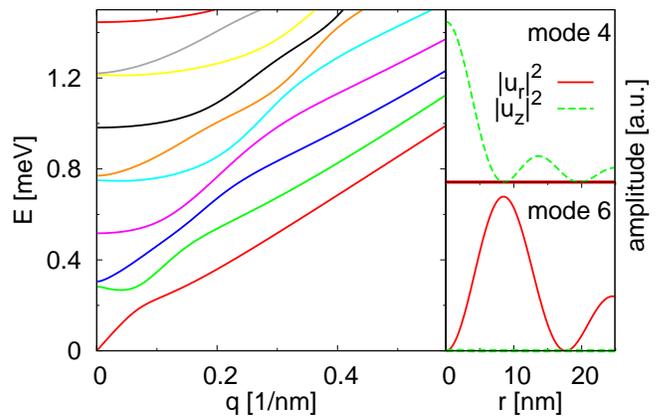}
\caption{(left) Dispersion relation and (right) two exemplary modes for $q$ = 0.002 nm$^{-1}$ of the one-dimensional wire phonons.}
\label{modesOmega}
\end{figure}

We consider a zinc-blende lattice in the (111) growth direction. The resulting lattice corresponds, in a nearest-neighbor approximation, to the wurtzite lattice \cite{Bykhovski:ApplPhysLett:96}. Calculations of strain in nanowires have shown that such a transformation is a good approximation \cite{Larsson:Nanotechnology:07}. The orientation is taken into account in the phonon modes via the velocities of sound which are now taken along the (111) direction. The dispersion relation as well as two exemplary modes are shown in Fig. \ref{modesOmega}.
\\

\noindent{\bf Quantum dot model and wire parameters}:
% \paragraph{Quantum dot model and wire parameters}
For the QD model, we take a spherically symmetric charge distribution with harmonic potential $\varphi_i({\bf x}) \propto \E^{-(r^2 + z^2)/2 a_i^2}$, with Gaussian radii $a_c$ = 5.8 nm and $a_v$ = 3.19 nm. For the wire, we use GaAs parameters: deformation potentials $D_c$ = -14.6 eV, $D_v$ = -4.8 eV \cite{Krummheuer:PhysRevB:02}, mass density $\rho_{\rm m}$ = 5370 kg/m$^3$, longitudinal and transverse sound velocities $v_l$ = 5400 m/s, $v_t$ = 2800 m/s, dielectric constant $\varepsilon_{\rm s}$ = 12.9, and wire radius $R$ = 25 nm. The piezoelectric constants for wurtzite GaAs are not known. We thus calculate the constants from the zinc-blende value $e_{14}$, again using the transformation described above \cite{Bykhovski:ApplPhysLett:96}. Using the value $e_{14} = -0.16$ Cm$^{-2}$ \cite{O::01}, we obtain an estimate for the wurtzite piezoelectric constants $e_{15} = e_{31} = 0.09$ Cm$^{-2}$ and $e_{33} = -0.18$ Cm$^{-2}$. We note that larger values for $e_{14}$ can be found in the literature \cite{Xin:ApplPhysLett:07} and that the transformation formula is an approximation, and thus the effect of the piezoelectric coupling could certainly be enhanced compared to the results shown here.

\section{Electron-phonon coupling mechanisms}\label{coup}

We consider two electron-phonon coupling mechanisms: the deformation potential and the piezoelectric coupling. In the following, we discuss and compare the coupling elements and the resulting phonon spectral densities. We restrict the discussions to the six energetically lowest phonon modes.\\

\begin{figure}[htb]
\includegraphics*[width=0.49\linewidth]{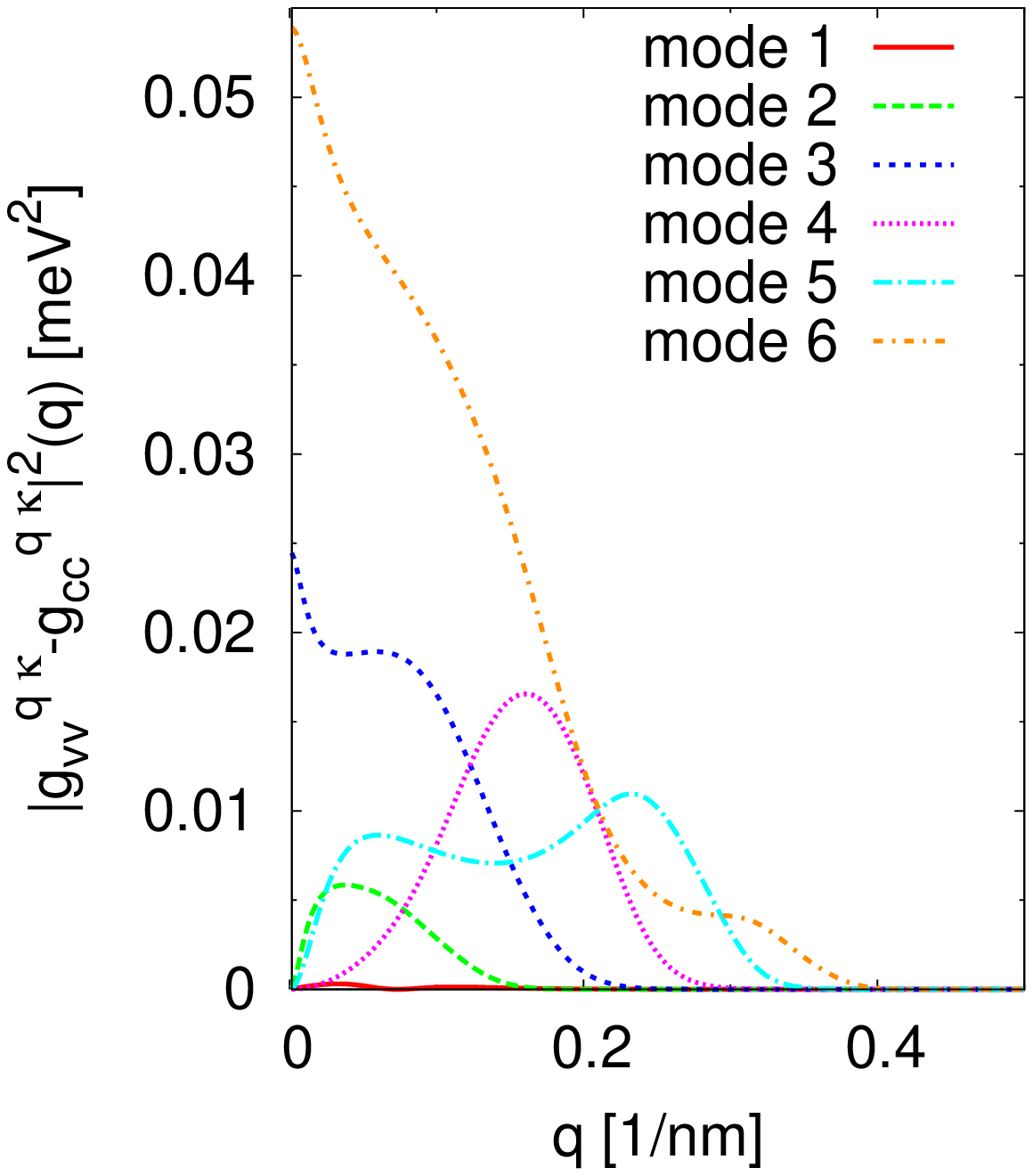}
\includegraphics*[width=0.49\linewidth]{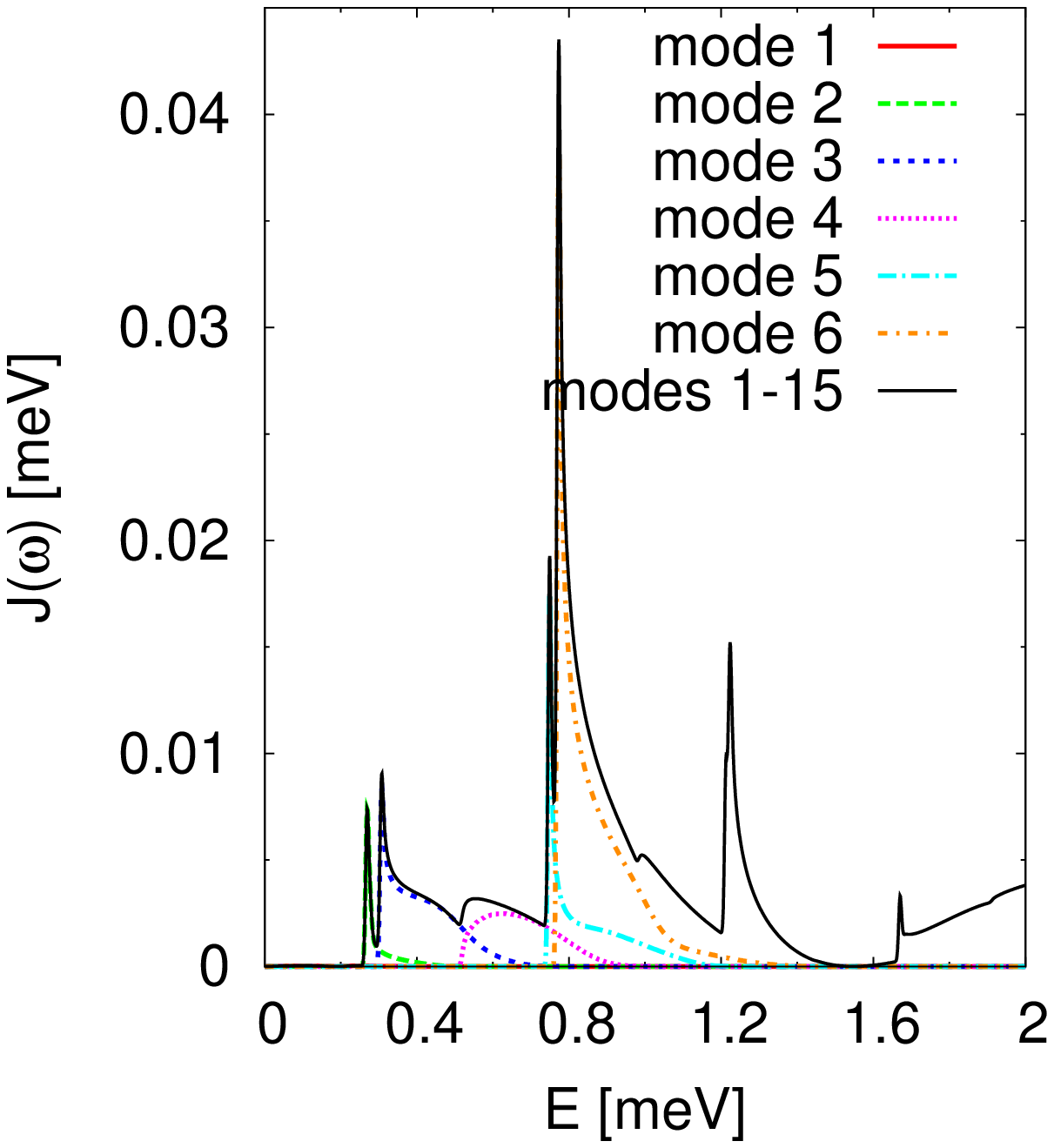}
\caption{(left) Coupling elements $|g_{v v}^{q \kappa} - g_{c c}^{q \kappa}|^2(q)$ ($L$ = 1 nm) and (right) phonon spectral density $J(\omega)$ for the deformation potential coupling for the energetically lowest phonon modes.}
\label{coup_DP}
\end{figure}

\noindent{\bf Deformation potential}:
% \paragraph{Deformation potential}
The coupling of the electronic system to the phonons via the deformation potential is just given by the divergence of the displacement \cite{Mahan::00}:
\begin{equation}\label{defPot}
V^i({\bf x}) = D_i \, \nabla \cdot {\bf u}({\bf x}),
\end{equation}
where $D_i$ is the deformation potential. The calculated coupling constants are shown in Fig. \ref{coup_DP}(left), while the corresponding spectral density is shown in Fig. \ref{coup_DP}(right). Modes 3 and 6 are almost purely radial for $q \approx 0$ (cf. Fig. \ref{modesOmega}), thus they couple strongly via the deformation potential. However, due to the density of states, the spectral density of mode 3 is relatively weak. The axial modes 4 and 5 only couple weakly via the deformation potential for $q \approx 0$ due to the restriction to diagonal strain elements $S_{i j} = (\partial_i u_j + \partial_j u_i)/2$ in Eq. (\ref{defPot}).
\\

\noindent{\bf Piezoelectric coupling}:
%\paragraph{Piezoelectric coupling}
For the piezoelectric coupling, the displacement strain causes an electrostatic potential which is determined via
\begin{equation}
\varepsilon_0 \varepsilon_{\rm s} \Delta \phi({\bf x}) = \nabla \cdot {\bf P}^{\rm pz},
\end{equation}
where $P^{\rm pz}_{k} = e_{k, ij} S_{i j}$ (piezoelectric constants $e_{k,ij}$) \cite{Mahan::73,Takagahara:PhysRevLett:93,Pokatilov:JApplPhys:04}. For the wurtzite structure, only three independent piezoelectric constants remain, leading to the polarization
\begin{eqnarray}\label{Ppz}
&&{\bf P}^{\rm pz}({\bf x}) = e_{1 5} \left( \partial_r u_z({\bf x}) + \partial_z u_r({\bf x}) \right) {\bf e}_r\\
&&\hspace*{0.5cm}+ \left\{ e_{31} \left( \partial_r u_r({\bf x}) + \frac{1}{r} u_r({\bf x}) \right) + e_{33} \partial_z u_z({\bf x}) \right\} {\bf e}_z. \nonumber
\end{eqnarray}
Assuming that the potential vanishes for $r \rightarrow \infty$ using $\varepsilon_{\rm s}$ = 1 outside the wire, we obtain an analytical solution for the electron-phonon potential $V({\bf x}) = -\E \phi({\bf x})$.

\begin{figure}[htb]
\includegraphics*[width=0.49\linewidth]{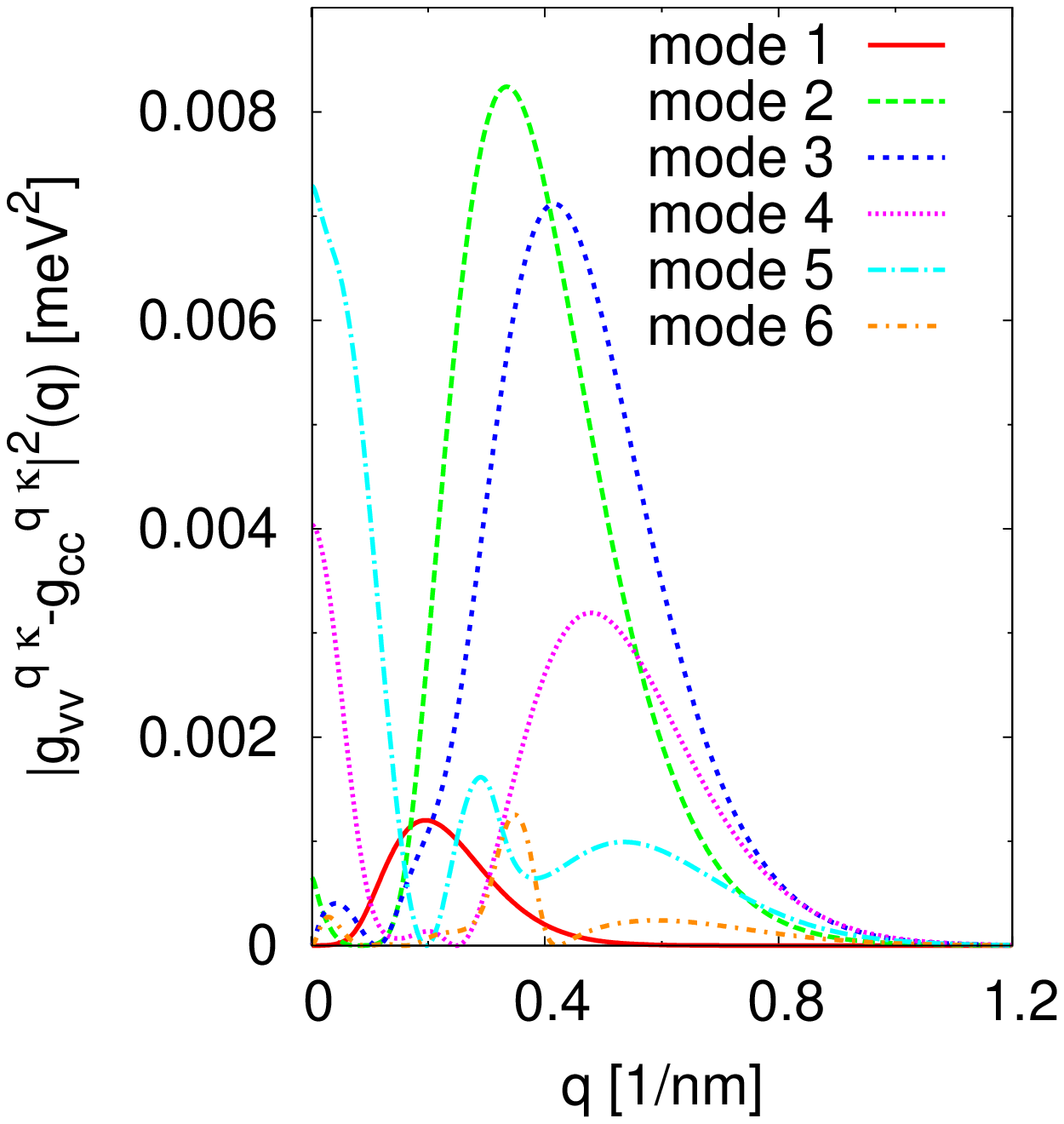}
\includegraphics*[width=0.49\linewidth]{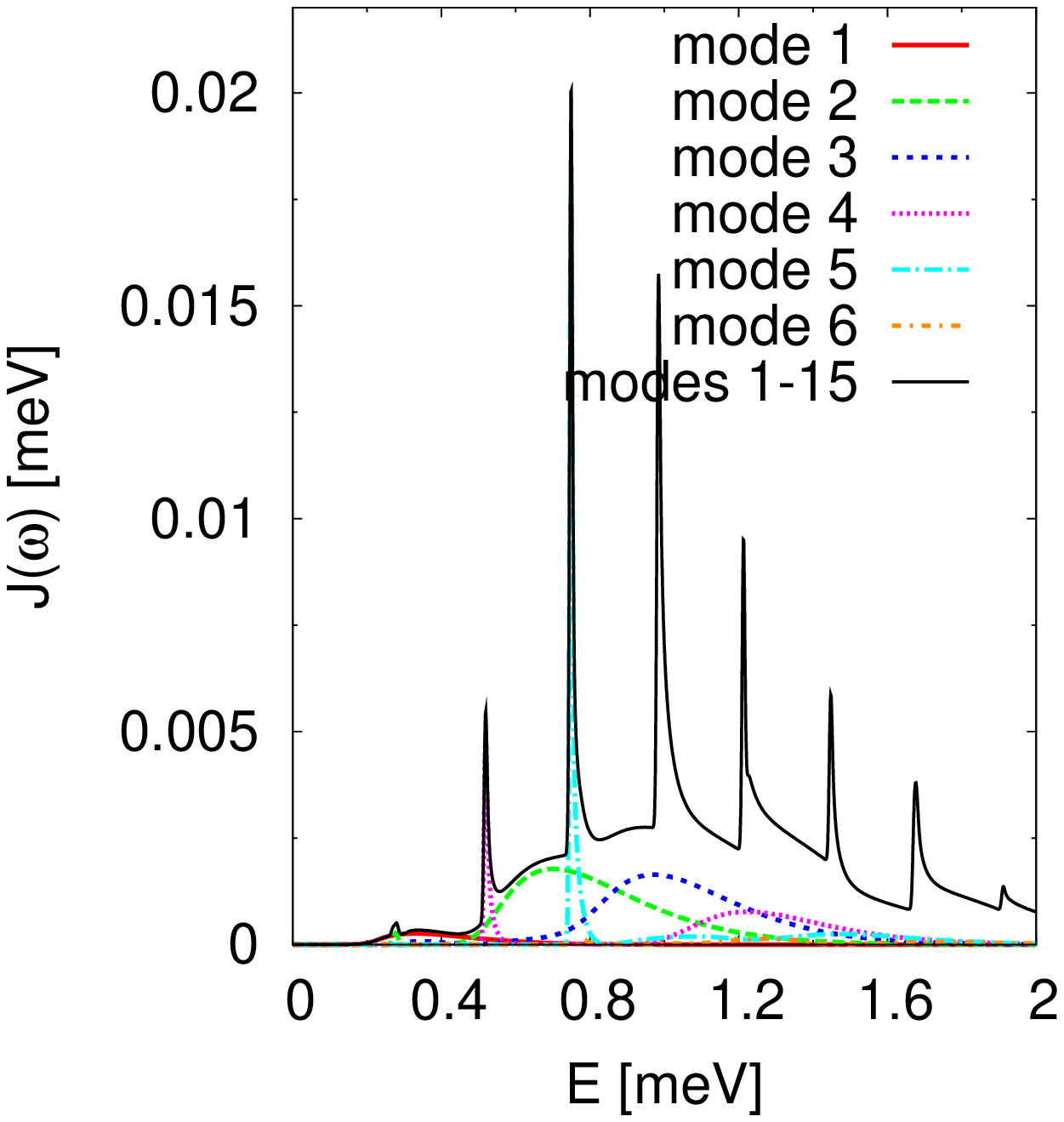}
\caption{(left) Coupling elements $|g_{v v}^{q \kappa} - g_{c c}^{q \kappa}|^2(q)$ ($L$ = 1 nm) and (right) phonon spectral density $J(\omega)$ for the piezoelectric coupling for the energetically lowest phonon modes.}
\label{coup_PZ}
\end{figure}

\begin{figure}[htb]
\includegraphics*[width=0.49\linewidth]{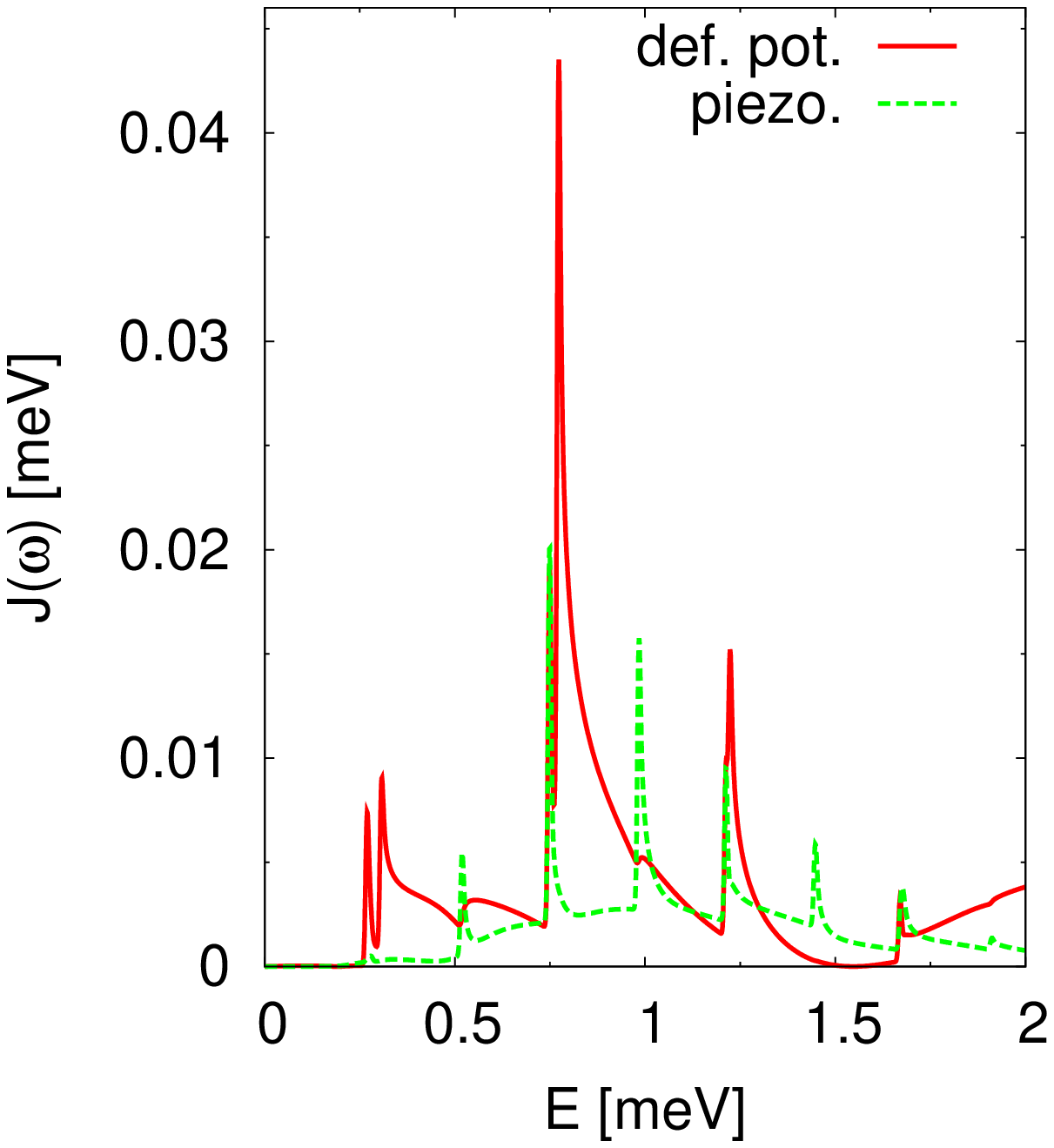}
\includegraphics*[width=0.49\linewidth]{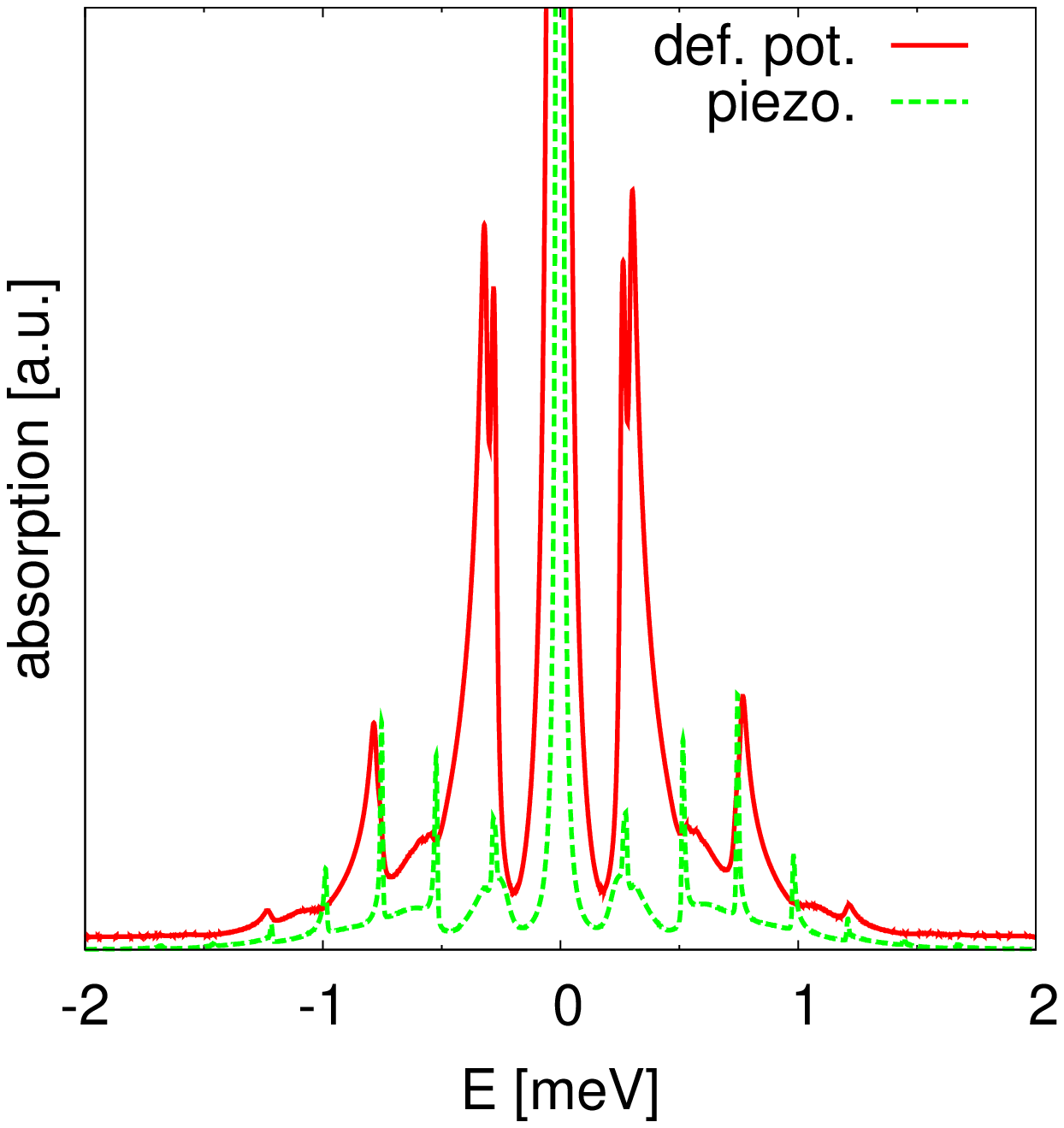}
\caption{Comparison of (left) the phonon spectral densities $J(\omega)$ and (right) the absorption spectra at $T$ = 77 K for the deformation potential and piezoelectric coupling.}
\label{cmp}
\end{figure}

In Figs. \ref{coup_PZ}(left) and \ref{coup_PZ}(right), the coupling elements and the corresponding spectral density is shown, respectively.
Compared to the deformation potential coupling, modes 4 and 5 couple strongly here for $q \approx 0$ since the piezoelectric potential [Eq. (\ref{Ppz})] takes into account the nondiagonal strain elements $S_{i j}$ which favor axial modes for small $q$ (cf. Fig. \ref{modesOmega}). This interplay between the deformation potential and the piezoelectric coupling can be seen nicely when comparing the spectral densities of the two coupling mechanisms, shown in Fig. \ref{cmp}(left).

\section{Absorption spectra}

\begin{figure}[htb]
\includegraphics*[width=0.49\linewidth]{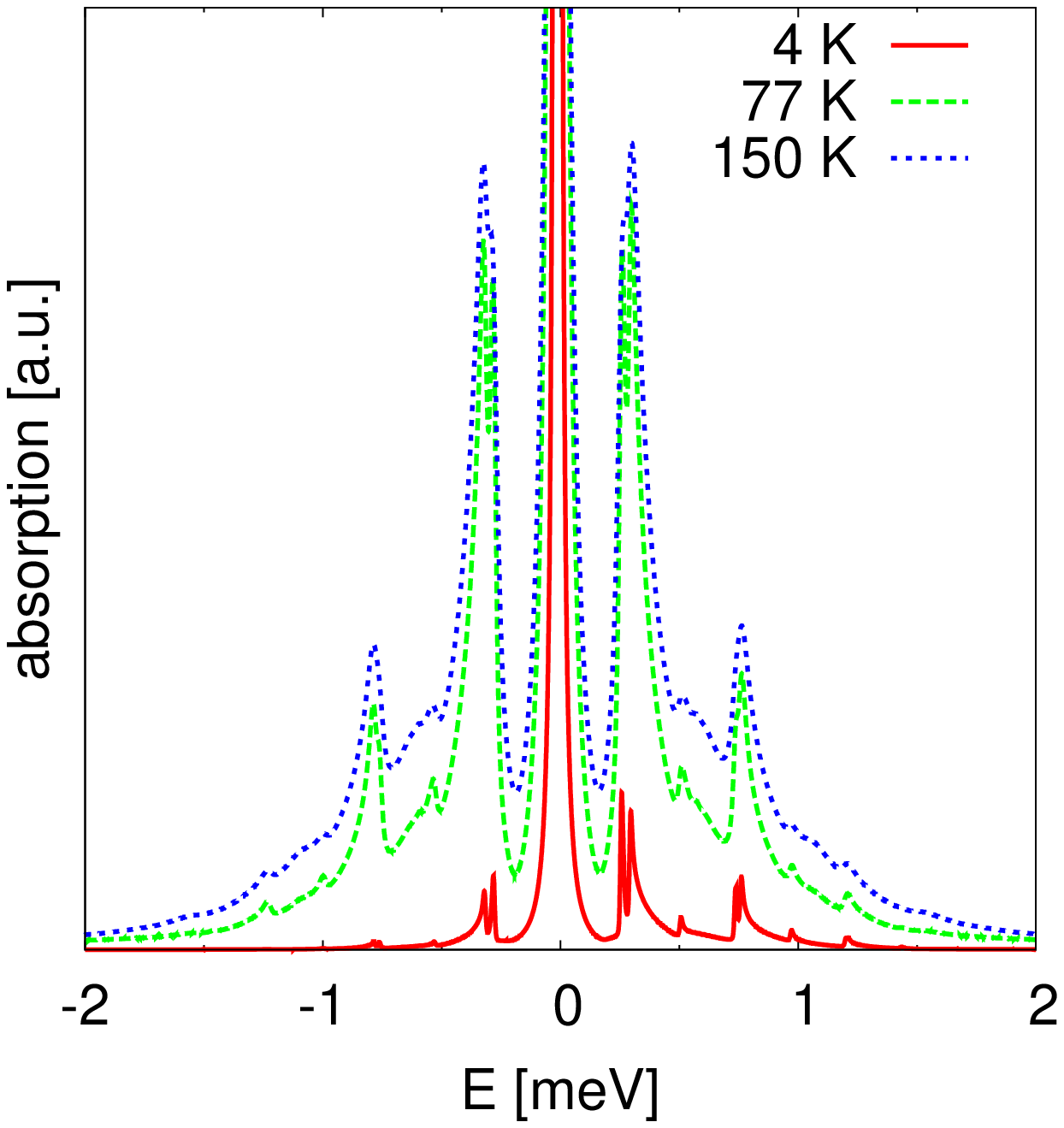}
\includegraphics*[width=0.49\linewidth]{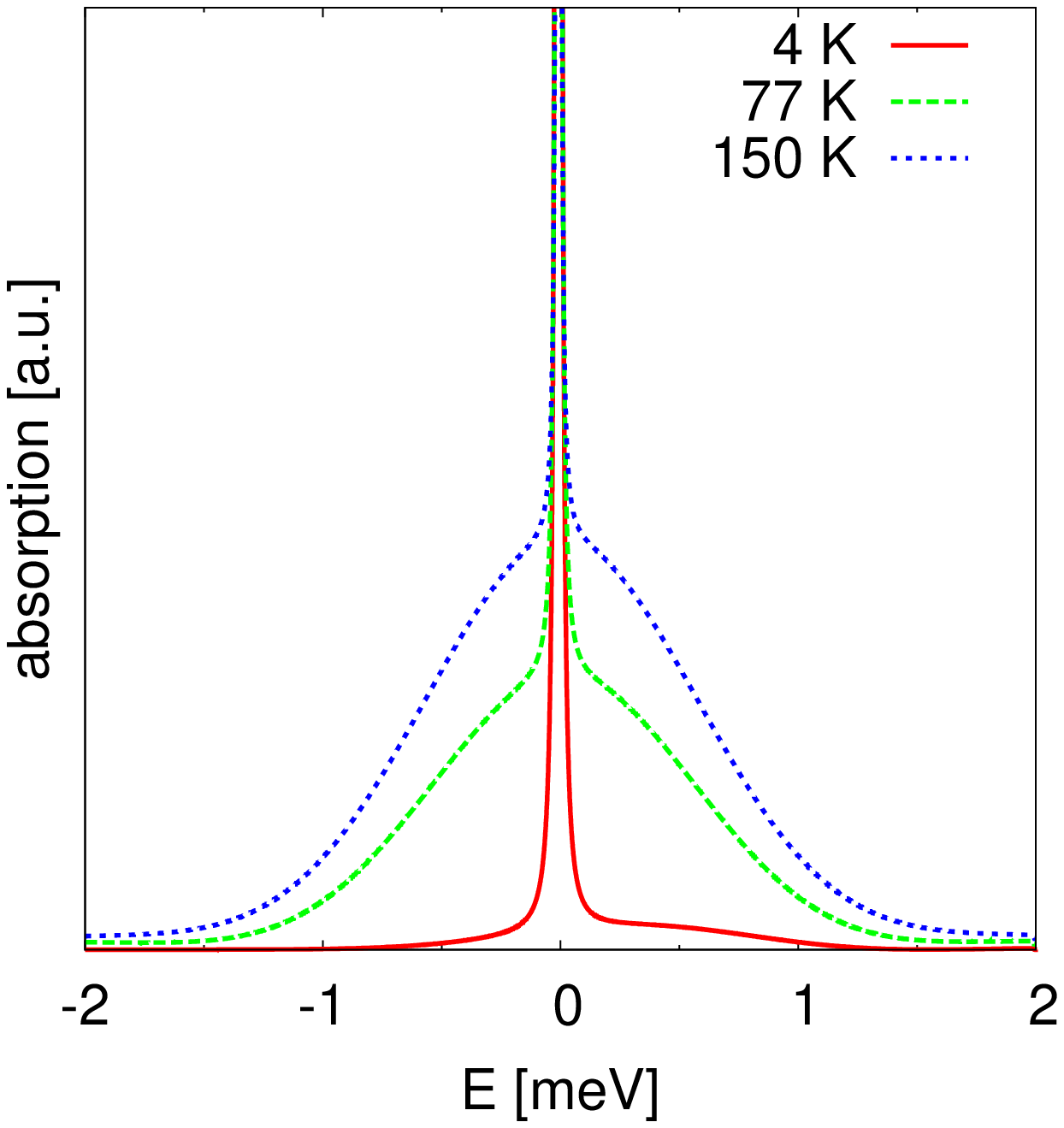}
\caption{Absorption spectrum of a QD interacting with (left) wire phonons via the deformation potential and the piezoelectric coupling and (right) bulk phonons via the deformation potential coupling for different temperatures.}
\label{abs}
\end{figure}

We now calculate the absorption within the IBM for the one-dimensional phonon modes. In Fig. \ref{abs}, the absorption spectrum for (left) a QD interacting with the wire modes is compared to (right) a QD exposed to bulk phonons for different temperatures. In both cases, a phenomenological (radiative) dephasing of $T_2 = 1 \, \mu$eV is used \cite{Borri:PhysRevLett:01}, and the system is excited resonantly on the polaron shifted transition energy $E = \hbar (\omega_{\rm gap} + \Delta)$. In Fig. \ref{cmp}(right), the different coupling mechanisms are compared in the absorption for the wire modes.
\\

\noindent{\bf Side peaks}:
% \paragraph{Side peaks}
Compared to the case of bulk phonons, where the linear dispersion of the acoustic phonon mode leads to continuous side bands attached to the ZPL, the discrete wire modes lead to a series of phonon side peaks. The side bands are strongly temperature dependent, where higher temperatures lead to both increased dephasing as well as a broader spectrum due to multi-phonon scattering processes. Furthermore, at low temperatures, the absorption/emission asymmetry can be seen. The peaks in the absorption spectra correspond to the extrema in the dispersion, where the density of states has a singularity, whereas their strength is determined by the magnitude of the coupling at the extrema \cite{Lindwall:PhysRevLett:07}. In addition to the radial modes 3 and 6 which couple strongly via the deformation potential, the axial modes 4 and 5 can be seen in the absorption which interact via the piezoelectric coupling. The dependence of the scattering mechanism on the mode character can be seen more clearly in Fig. \ref{cmp}(right), where the two mechanisms are compared.\\

\noindent{\bf ZPL broadening}:
% \paragraph{ZPL broadening}
As was shown in Ref. \cite{Lindwall:PhysRevLett:07}, the phonon mode extending to vanishing energy leads to a temperature dependent broadening of the ZPL. This can be seen nicely in Fig. \ref{abs}(left) for increasing temperature. For the piezoelectric coupling, no considerable contribution to the ZPL broadening is found due to the form of the piezoelectric potential for $q \rightarrow 0$.

\section{Conclusion}

We have studied the interaction of a quantum dot in a catalytically grown GaAs nanowire via the deformation potential and piezoelectric electron-phonon coupling. In general, the electron-phonon interaction leads to discrete side peaks, which are strongly temperature dependent, and a broadening of the zero-phonon line. Since the deformation potential couples strongly only to radial modes, the piezoelectric coupling can become important for modes which are mainly axial and thus can play an important role for the determination of the optical spectrum. The inclusion of the piezoelectric coupling does not lead to an appreciable contribution to the zero-phonon line broadening. It should be mentioned that larger values of the piezoelectric constants than used here have been reported in the literature, and thus the influence of the piezoelectric coupling could become more important. Furthermore, the piezoelectric coupling is strongly dependent on the quantum dot geometry, and thus the relative as well as the absolute strengths of the coupling constants must be considered for each case.

As an outlook, the size and shape dependence of the piezoelectric coupling to the one-dimensional wire phonon modes must be further investigated. In addition, the torsional modes, which are neglected in this investigation, should be considered to ascertain their interaction strength via the piezoelectric coupling.

\begin{acknowledgments}
We would like to thank the Swedish Research Council (VR) for financial support.
\end{acknowledgments}

%\bibliographystyle{pss}
%\bibliography{quWire}

\begin{thebibliography}{[10]}

\bibitem{Bjork:NanoLett:02}% article
 \textsc{M.\,T. Bj{\"o}rk},  \textsc{B.\,J. Ohlsson},  \textsc{T.~Sass},
  \textsc{A.\,I. Persson},  \textsc{C.~Thelander},  \textsc{M.\,H. Magnusson},
  \textsc{K.~Deppert},  \textsc{L.\,R. Wallenberg},  and
  \textsc{L.~Samuelson}\iffalse One-dimensional steeplechase for electron
  realized\fi,
 \jr{Nano Lett.} \textbf{2}, 87 (2002).


\bibitem{Agarwal:ApplPhysA:06}% article
 \textsc{R.~Agarwal} and  \textsc{C.\,M. Lieber}\iffalse Semiconductor
  nanowires: optics and optoelectronics\fi,
 \jr{Appl. Phys. A} \textbf{85}, 209 (2006).


% \othercit
% \bibitem{Nielsen::00}% book
%  \textsc{M.\,A. Nielsen} and  \textsc{I.\,L. Chuang},
% Quantum Computation and Quantum Information (Cambridge University Press,
%   Cambridge, 2000).


\bibitem{Moreau:PhysRevLett:01}% article
 \textsc{E.~Moreau},  \textsc{I.~Robert},  \textsc{L.~Manin},
  \textsc{V.~Thierry-Mieg},  \textsc{J.\,M. G\'erard},  and
  \textsc{I.~Abram}\iffalse Quantum cascade of photons in semiconductor quantum
  dots\fi,
 \jr{Phys. Rev. Lett.} \textbf{87}, 183601 (2001).


\bibitem{Kako:NatureMater:06}% article
 \textsc{S.~Kako},  \textsc{C.~Santori},  \textsc{K.~Hoshino},
  \textsc{S.~G\"otzinger},  \textsc{Y.~Yamamoto},  and
  \textsc{Y.~Arakawa}\iffalse A gallium nitride single-photon source operating
  at 200 {K}\fi,
 \jr{Nature Mater.} \textbf{5}, 887 (2006).


% \bibitem{Bimberg:JPhysD:05}% article
%  \textsc{D.~Bimberg}\iffalse Quantum dots for lasers, amplifiers and
%   computing\fi,
%  \jr{J. Phys. D} \textbf{38}, 2055 (2005).


\bibitem{Borgstroem:NanoLett:05}% article
 \textsc{M.~Borgstr\"om},  \textsc{V.~Zwiller},  \textsc{E.~M\"uller},  and
  \textsc{A.~Imamoglu},
 \jr{Nano Lett.} \textbf{5}, 1439 (2005).


% \bibitem{Chow:IEEEJQuantumElectron:05}% article
%  \textsc{W.\,W. Chow} and  \textsc{S.\,W. Koch}\iffalse Theory of semiconductor
%   quantum-dot laser dynamics\fi,
%  \jr{IEEE J. Quantum Electron.} \textbf{41}, 495 (2005).


\bibitem{Lindwall:PhysRevLett:07}% article
 \textsc{G.~Lindwall},  \textsc{A.~Wacker},  \textsc{C.~Weber},  and
  \textsc{A.~Knorr}\iffalse Zero-phonon linewidth and phonon satellites in the
  optical absorption of nanowire-based quantum dots\fi,
 \jr{Phys. Rev. Lett.} \textbf{99}, 087401 (2007).


\bibitem{Borri:PhysRevLett:01}% article
 \textsc{P.~Borri},  \textsc{W.~Langbein},  \textsc{S.~Schneider},
  \textsc{U.~Woggon},  \textsc{R.\,L. Sellin},  \textsc{D.~Ouyang},  and
  \textsc{D.~Bimberg}\iffalse Ultralong dephasing time in {I}n{G}a{A}s quantum
  dots\fi,
 \jr{Phys. Rev. Lett.} \textbf{87}, 157401 (2001).

\bibitem{Krummheuer:PhysRevB:02}% article
 \textsc{B.~Krummheuer},  \textsc{V.\,M. Axt},  and  \textsc{T.~Kuhn}\iffalse
  Theory of pure dephasing and the resulting absorption line shape in
  semiconductor quantum dots\fi,
 \jr{Phys. Rev. B} \textbf{65}, 195313 (2002).


\bibitem{Forstner:PhysStatusSolidiB:02}% article
 \textsc{J.~F\"orstner},  \textsc{K.\,J. Ahn},  \textsc{J.~Danckwerts},
  \textsc{M.~Schaarschmidt},  \textsc{I.~Waldm\"uller},  \textsc{C.~Weber},
  and  \textsc{A.~Knorr}\iffalse Light propagation- and many-particle-induced
  non-{L}orentzian lineshapes in semiconductor nanooptics\fi,
 \jr{Phys. Status Solidi B} \textbf{234}, 155 (2002).


\bibitem{Forstner:PhysRevLett:03}% article
 \textsc{J.~F\"orstner},  \textsc{C.~Weber},  \textsc{J.~Danckwerts},  and
  \textsc{A.~Knorr}\iffalse Phonon-assisted damping of {R}abi oscillations in
  semiconductor quantum dots\fi,
 \jr{Phys. Rev. Lett.} \textbf{91}, 127401 (2003).


\othercit
\bibitem{Zimmermann::02}% inproceedings
 \textsc{R.~Zimmermann} and  \textsc{E.~Runge},
 in: Proc. 26th ICPS Edinburgh, A. R. Long and J. H. Davies (eds.), IOP Conf. Series,  Vol.\,171 (IOP Publishing,
  Bristol, 2002),
paper M 3.1.


\bibitem{Muljarov:PhysRevLett:05}% article
 \textsc{E.\,A. Muljarov},  \textsc{T.~Takagahara},  and
  \textsc{R.~Zimmermann}\iffalse Phonon-induced exciton dephasing in quantum
  dot molecules\fi,
 \jr{Phys. Rev. Lett.} \textbf{95}, 177405 (2005).


\bibitem{Machnikowski:PhysRevLett:06}% article
 \textsc{P.~Machnikowski}\iffalse Change of decoherence scenario and appearance
  of localization due to reservoir anharmonicity\fi,
 \jr{Phys. Rev. Lett.} \textbf{96}, 140405 (2006).


\bibitem{Muljarov:PhysRevLett:07}% article
 \textsc{E.\,A. Muljarov} and  \textsc{R.~Zimmermann}\iffalse Exciton dephasing
  in quantum dots due to {LO}-phonon coupling: An exactly solvable model\fi,
 \jr{Phys. Rev. Lett.} \textbf{98}, 187401 (2007).


\othercit
\bibitem{Mahan::00}% book
 \textsc{G.\,D. Mahan},
Many-Particle Physics (Plenum Publishers, New York, 2000).


\bibitem{Stroscio:JApplPhys:94}% article
 \textsc{M.\,A. Stroscio},  \textsc{K.\,W. Kim},  \textsc{S.~Yu},  and
  \textsc{A.~Ballato}\iffalse Quantized acoustic phonon modes in quantum wires
  and quantum dots\fi,
 \jr{J. Appl. Phys.} \textbf{76}, 4670 (1994).


\bibitem{Bykhovski:ApplPhysLett:96}% article
 \textsc{A.\,D. Bykhovski},  \textsc{V.\,V. Kaminski},  \textsc{M.\,S. Shur},
  \textsc{Q.\,C. Chen},  and  \textsc{M.\,A. Khan}\iffalse Piezoresistive
  effect in wurtzite n-type gan\fi,
 \jr{Appl. Phys. Lett.} \textbf{68}, 818 (1996).


\bibitem{Larsson:Nanotechnology:07}% article
 \textsc{M.\,W. Larsson},  \textsc{J.\,B. Wagner},  \textsc{M.~Wallin},
  \textsc{P.~H{\aa}kansson},  \textsc{L.\,E. Fr\"oberg},
  \textsc{L.~Samuelson},  and  \textsc{L.\,R. Wallenberg}\iffalse Strain
  mapping in free-standing heterostructured wurtzite inas/inp nanowires\fi,
 \jr{Nanotechnology} \textbf{18}, 015504 (2007).


\othercit
\bibitem{O::01}% book
 \textsc{O.~Madelung},  \textsc{U.~R\"ossler},  and  \textsc{M.~Schulz} (eds.),
Group IV Elements, IV-IV and III-V Compounds, Landolt-B\"ornstein - Group III,
  Vol.\,41A1a (Springer, Berlin, 2001).


\bibitem{Xin:ApplPhysLett:07}% article
 \textsc{J.~Xin},  \textsc{Y.~Zheng},  and  \textsc{E.~Shi}\iffalse
  Piezoelectricity of zinc-blende and wurtzite structure binary compounds\fi,
 \jr{Appl. Phys. Lett.} \textbf{91}, 112902 (2007).


\othercit
\bibitem{Mahan::73}% inbook
 \textsc{G.\,D. Mahan},
Polarons in Ionic Crystals and Polar Semiconductors,
 (North-Holland, Amsterdam, 1973), chap. Polarons in heavily doped
  semiconductors,  p.\,553.


\bibitem{Takagahara:PhysRevLett:93}% article
 \textsc{T.~Takagahara}\iffalse Electron-phonon interactions and excitonic
  dephasing in semiconductor nanocrystals\fi,
 \jr{Phys. Rev. Lett.} \textbf{71}, 3577 (1993).

\bibitem{Pokatilov:JApplPhys:04}% article
 \textsc{E.~P.~Pokatilov}, \textsc{D.~L.~Nika}, and \textsc{A.~A.~Balandin}\iffalse Confined electron-confined phonon scattering rates in wurtzite AlN/GaN/AlN heterostructures\fi,
 \jr{J. Appl. Phys.} \textbf{95}, 5626 (2004).

\end{thebibliography}

% The resulting bibliography-output (the contents of the .bbl file)
% must be pasted back into this file before submission.
%
% Replace the following example bibliography with your references
% before submission:
%

\providecommand{\WileyBibTextsc}{}
\let\textsc\WileyBibTextsc
\providecommand{\othercit}{}
\providecommand{\jr}[1]{#1}
\providecommand{\etal}{~et~al.}

\end{document}